\documentclass[conference]{IEEEtran}
\IEEEoverridecommandlockouts
\usepackage{cite}
\usepackage{amsmath,amssymb,amsfonts}
\usepackage{algorithmic}
\usepackage{graphicx}
\usepackage{textcomp}
\usepackage{xcolor}
\usepackage[utf8]{inputenc}
\usepackage[english]{babel}
\usepackage{enumitem}
\usepackage{hyperref}

\begin{document}

\title{A Private Smart Wallet \\ with Probabilistic Compliance\\

}


\author{
    \IEEEauthorblockN{
        Andrea Rizzini,
        Marco Esposito,
        Francesco Bruschi,
        Donatella Sciuto
    }
    \IEEEauthorblockA{
        Dipartimento di Elettronica, Informazione e Bioingegneria (DEIB),\\
        Politecnico di Milano, Milano, Italy\\
        Email: \{andrea.rizzini, marco.esposito, francesco.bruschi, donatella.sciuto\}@polimi.it
    }
}

\maketitle

\begin{abstract}
We propose a privacy-preserving smart wallet with a novel invitation-based private onboarding mechanism. The solution integrates two levels of compliance in concert with an authority party: a proof of innocence mechanism and an ancestral commitment tracking system using bloom filters for probabilistic UTXO chain states. Performance analysis demonstrates practical efficiency: private transfers with compliance checks complete within seconds on a consumer-grade laptop, and overall with proof generation remaining low. On-chain costs stay minimal, ensuring affordability for all operations on Base layer 2 network. The wallet facilitates private contact list management through encrypted data blobs while maintaining transaction unlinkability. Our evaluation validates the approach's viability for privacy-preserving, compliance-aware digital payments with minimized computational and financial overhead.
\end{abstract}


\section{Introduction}
With over 300 million people globally using blockchain and the market projected to reach \$32.69 billion by 2024 \cite{kumar2024blockchain}, adoption is likely to become as widespread as online banking today. However, as this user base grows, the incidence of scams and malicious uses will also rise, highlighting the need for tools enforcing compliance to prevent misuse. Blockchain technically enables universal traceability, but this conflicts with the fundamental right to privacy—without which mainstream adoption remains unlikely. Given that cryptocurrency enables pseudonymous, borderless value transfer with near-instant settlement finality, robust Anti-Money Laundering (AML) mechanisms become critical to prevent the technology from enabling large-scale financial crimes. The urgency of cryptocurrency AML measures was highlighted when OFAC sanctioned Tornado Cash for processing \$7 billion in transactions since 2019, including \$455 million from North Korea's Lazarus Group and over \$100 million from the 2022 Harmony Bridge and Nomad protocol exploits \cite{usd2022}. Several techniques to launder stolen funds have since been at play, including multiple hierarchical layers of transactions to obscure the money trail \cite{wu2023towards}. In parallel, reconciling blockchain's transparency with GDPR's privacy mandates and data subject rights presents unique technical and compliance challenges for adoption \cite{schellinger2022yes}. Fortunately, programmable cryptography offers promising tools to enforce privacy-preserving compliance rules ex-ante, potentially avoiding costly post-facto litigation and enforcement actions \cite{bamberger2022verification}. 

\subsubsection*{\text{Our contributions}} 
We implement a mixer-based wallet application with privacy-preserving onboarding and transfers. Additionally, we integrate a two-tiered compliance framework: a proof of innocence mechanism for direct or transitive checks, and a novel ancestral commitment tracking system using bloom filters for probabilistic UTXO chain state propagation during internal transfers. From a technical standpoint, our solution addresses a limitation in existing privacy pools: it eliminates the need to explicitly forward transaction history between users while maintaining verifiable compliance across multiple coin merges \cite{buterin2024blockchain}. Unlike sequential proof approaches that rely on passing secret information forward, our solution enables recipients to independently verify their funds source legitimacy  without revealing transaction linkages. 

Sections are organized as follows: Section II describes the state of the art, Section III presents our solution, Section IV presents the experimental results along with cost analyses, and Section V concludes.

\section{State of the art}

\subsection{Transaction confidentiality} In a blockchain token transaction, two primary entities are involved: a sender and a receiver, which may represent accounts or smart contracts. While observers on the blockchain may not be able to directly identify the individuals controlling these addresses, they can still view all transaction details in plaintext, including transferred amounts, deposits, withdrawals, and calldata. The key is that link between sender and receiver is preserved at all times allowing the exercise of tracking tools since the early days of the industry \cite{spagnuolo2014bitiodine}. To address this exposure, specialized mechanisms have been developed to obscure the link between sender and receiver.

\textit{Stealth Addresses.} An early practice to preserve privacy involved generating a new address for each transaction receipt. This required significant user interaction and coordination between parties. With stealth addresses users could generate a unique, one-time address for each transaction, or dual-key pairs to enable non-interactive generation of unlinkable addresses \cite{wahrstatter2024basesap}. Notably, stealth addresses do not obscure transaction flows, and funds remain traceable at pseudonymous endpoints.

\begin{figure}
    \centering
    \includegraphics[width=0.85\linewidth]{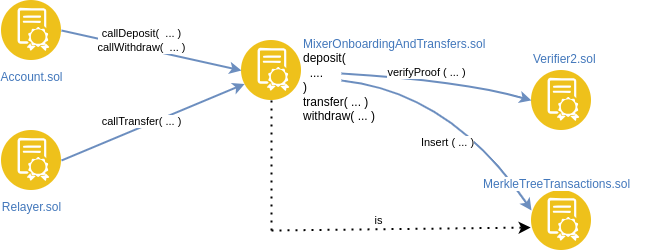}
    \caption{Contracts interactions, example for the deposit method.}
    \label{fig:ContractsInteraction}
\end{figure}

\textit{Mixers.} These are objects designed to obscure the link between sender and receiver by pooling and mixing users' assets. This process is meant to conceals transaction paths, making it difficult to trace specific assets back to their origin. A mixer is primarily implemented as a non-custodial protocol, although some trusted third-party implementations exist \cite{barbereau2023beyond}. Most common non-custodial mixers maintain two primary data structures at the smart contract level: 
\begin{itemize}
    \item R, an append-only list of commitments typically organized as a Merkle tree
    \item N, an append-only list of nullifiers, which serve as unique identifiers for each withdrawal.
\end{itemize}
Often, relayers are employed to avoid to prevent de-anonymization, as they submit transactions on behalf of users without holding custody of the assets. For example, in Tornado Cash Nova, users can choose to use a relayer for internal transfers; without it, the sender’s identity is exposed. In our solution, relayer use is mandatory, with fees covered by a Paymaster contract instead of the user (see Fig. \ref{fig:ContractsInteraction}). 

The effectiveness of a mixer depends on its anonymity measure \( A = f(n, t, d) \), where \( n \) represents the number of participants in the mixing pool, \( t \) denotes the time interval between deposit and withdrawal operations, and \( d \) represents the distribution of transaction values.
In the deposit process, a user generates a random secret \( s \) and computes a commitment \( L = h(s + 1) \), where \( h \) is a cryptographic hash function. The user then sends their assets along with \( L \) to the mixer's smart contract, which adds \( L \) to the commitment set \( R \).
During withdrawal, the user computes a nullifier \( U = h(s + 2) \). The user then generates a zero-knowledge proof \( \pi \), a method for proving knowledge of data without revealing it \cite{9520375}, demonstrating both that they know a secret \( s \) corresponding to some commitment in \( R \) and that \( U \) is the correct nullifier for that secret. The contract verifies \( \pi \) and checks that \( U \notin N \) to prevent double-spending. If valid, the state transition is confirmed as the nullifier \( U \) is added to \( N \) and the funds are released.

Other mixers for UTXO-based blockchains like CoinJoin provide similar privacy benefits through different mechanisms, such as multiple transactions batching, combining inputs and outputs typically in fixed-sized amount, in ways that obscure the transaction graph \cite{maxwell2013coinjoin}.

\subsection{UTXO Models with arbitrary denominations for account-based blockchains}
\label{sec:UTXO_Models}
A UTXO graph represents the flow of funds through individual transactions. Each UTXO acts as a node in the graph, and each transaction connecting UTXOs forms the edges between nodes. Unlike UTXO-based blockchains where state is tracked
through a graph of unspent outputs, account-based blockchains maintain state through account balances. UTXO systems have certain algebraic properties, like a Church-Rosser property, that distinguish them from account-based systems and ensure that double-spending is impossible within the system \cite{gabbay2021algebras}. In account-based systems, like Ethereum and EVM-based chains, transaction validity can depend on global state that may be modified by intervening transactions.
However, it is possible to combine the account model's global state with UTXO-style transaction validation rules. A smart contract can maintain an internal UTXO set as part of its state, where each UTXO is represented as a data structure containing value and ownership conditions. Transaction ordering and state updates still follow account-based semantics, while the contract logic enforces algebraic properties of native UTXO systems like local validation and transaction commutativity. In the context of mixers, the contract allows arbitrary denomination by splitting and merging through JoinSplit transactions that maintain privacy \cite{jivanyan2019lelantus}. Each commitment contains both a value and blinding factor, allowing users to prove in zero-knowledge that: i. They have authority over the input UTXOs; ii. The sum of input values equals the sum of output values; iii. All output values are non-negative.
Blockchains like Zcash natively implements JoinSplit transactions to ensure privacy. Instead, Tornado Cash Nova and Railgun are privacy-preserving platforms that utilize the UTXO model with JoinSplit transactions, but implement them differently.  Nova represents each UTXO with three components (amount, public key, blinding), where the commitment is computed as a hash of these values and the nullifier incorporates the merkle path and a signature \cite{tornado_cash_docs}. This extends the basic commitment scheme used in earlier versions like Tornado Cash Core. Railgun similarly uses encrypted notes (UTXOs) containing public key, amount, token ID and a randomness field, but generates nullifiers deterministically by combining the spending key with merkle path indices, ensuring unique nullification \cite{railgun_wiki}. Our system adopts the JoinSplit model used in Tornado Cash Nova; specifically, we support joinsplits with either 2 or 16 input notes and always exactly 2 outputs. This design aligns with the constraints of zk-SNARK circuits implemented in Circom, which cannot handle dynamic input sizes. 

\subsection{On-chain regulatory compliance}

Privacy-preserving protocols may implement different solutions to fully comply with AML/CFT requirements \cite{burleson2022privacy}, or at least mitigate the risk of law enforcement action. Below is a preliminary classification of the utility of these solutions.

\subsubsection*{\textbf{Selective de-anonymization}} This strategy can be implemented either voluntarily, for personal fiscal accountability, or upon request. While Tornado Cash introduced a basic opt-in reporting mechanism through note presentation (i.e. hash of the deposit’s secret and nullifier), this approach had a significant drawback: once revealed, the transaction link would become permanently public. Recent protocols have overcome this limitation with advanced selective disclosure mechanisms \cite{railgun_wiki}. For instance, users can generate time-limited reports proving transaction integrity or provide viewing keys for access while preserving privacy from the broader network. However, using a viewing key is a drastic form of selective disclosure, as it lets a trusted third party decrypt and verify all transactions. Protocols like Railgun support optional viewing key functionality, though not as a requirement. This binary disclosure problem persists even with multiple viewing keys: if Alice has made 10 private transactions and a compliance team requests verification, she can choose which transactions to reveal (by providing their respective viewing keys), but for each transaction revealed, she must disclose its complete details.
To tackle the counter-party risk problem through cryptographic guarantees rather than trust assumptions, one can leverage the general-purpose nature of zkSNARKs to create bilateral proofs \cite{buterin2024blockchain}: Alice can generate a proof to show that either (1) her withdrawal belongs to a compliant set of commitments, (2) she is the centralized exchange (CEX) compliance team, or (3) more than $t$ seconds have passed. The actual CEX, receiving the proof in real-time, can conclude the first statement must be true because they know they didn't create the proof themselves and the timestamp is fresh. If the proof leaks, others can verify it but they cannot determine which of the three conditions made it valid. More advanced solutions, like Derecho \cite{beal2023derecho}, add an optional disclosure layer with proof-carrying data for selective attestation of fund origins. 

\subsubsection*{\textbf{Prevention and deterrence}} \label{sec:prevention} Screening mechanisms can be placed at either entry points, exit points, or both within the protocol. Specifically, blocking or funds-locking policies can be enacted by the protocol's relayer entities upon deposit and/or when users request withdrawals from the protocol. An example of this is seen in wallets interacting with the Railgun protocol \cite{railgun_2024}, where sanctioned lists from regulatory bodies like OFAC's SDN list \cite{ofac_sdn_list_2024} are used as input data to restrict addresses on or associated with these lists from joining the anonymity set \cite{popov2024blockchain}. Cross-referencing primarily relies on transitivity check operations of varying computational cost, typically conducted by on-chain intelligence services such as TRM Labs or Chainalysis. A transitivity check examines the connection depth between addresses: given a source address $s$ and a target address $t$, it determines if there exists a path of length $k \leq n$ where $n$ is the maximum hop count, such that $s \rightarrow^k t$ through intermediary addresses. For a given address $v$, its $n$-hop transaction neighborhood is defined as:

$N_n(v) = \{u \mid \exists \text{ path of length } k \leq n \text{ from } v \text{ to } u\}$ The computational complexity grows exponentially with \( n \) as \( O(d^n) \), where \( d \) is the average number of transaction partners per address, showing a major limit of these services. To mitigate potential vulnerabilities during the computation period, most privacy pool protocols adopt an "inclusion with delay" or "unshield-only standby period" (following terminology from Railgun which sets it to 60 minutes).

\subsubsection*{\textbf{Self-regulatory restrictions}}
Beyond preventive measures, protocols may implement restrictive compliance mechanisms through privacy-preserving KYC solutions. For example, the Hinkal protocol is based on KYC attestations from established providers like Coinbase and Binance, implemented through soul-bound tokens \cite{hinkal_docs}. These non-transferable tokens can serve as on-chain credentials to enforce tiered access levels \cite{ohlhaver2022decentralized}. Additionally, transactional limits and geo-restrictions can be implemented at the application level to further enhance compliance measures.

\section{Proposed solution}

\begin{figure}
    \centering
    \includegraphics[width=0.85\linewidth]{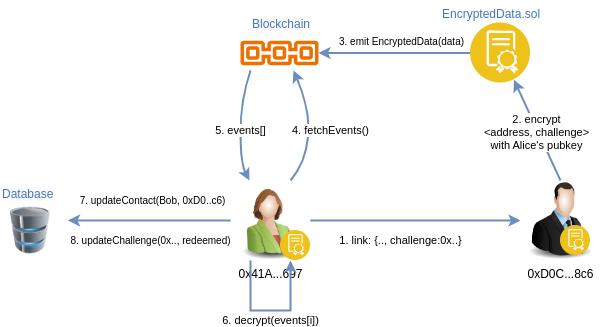}
    \caption{How Alice synchronizes with Bob’s contact.}
    \label{fig:Synch}
\end{figure}

This paper proposes an EVM-compatible wallet that combines standard functions with a built-in privacy layer and an improved onboarding experience, designed to be as intuitive as modern digital payment apps. Furthermore, we describe a new double-layered compliance scheme designed to mitigate the risks of illicit on-chain activities. The Joint Split transaction scheme implemented in Tornado Cash Nova and audited \cite{gulamov2021tornado}
provides the foundational UTXO model and zero-knowledge proof system, implemented via Circom for circuit arithmetization and Snarkjs for the Groth16 proof system, that our private wallet builds upon (see Section~\ref{sec:UTXO_Models}).
Our solution implements ERC-4337 account abstraction \cite{erc4337_architecture_2024}, where each user is assigned a smart contract account that serves as their wallet. This account contract is deployed either when a user registers with an invitation code or when they are onboarded by an existing user. The account contract implements core functionality including, registration in the protocol's pool, user operation validation and posting encrypted data blobs (via insertIntoEncryptedData()) that the application layer can use for contact synchronization (Fig. \ref{fig:Synch}). 

\subsection{Deposit with Bootstrapping}
\label{sec:deposit}
During the first deposit, a bootstrapping process is carried out by an authority through a three-step protocol involving the depositor, the authority, and the mixer acting as an intermediary. Subsequent deposits reference the initial state established during this process.

\subsubsection{Request bootstrapping by the depositor (D)}
\begin{itemize}
    \item \( \text{UTXO} = \{ ...\} \), \( C = \text{hash}(\text{UTXO}) \)
    \item \(\pi_D \): proving the expected fields are included in the UTXO and the hash computation
    \item Send \(\{C, \pi_D, pk_D \}\) to the mixer
    \item The mixer checks wether this is D's first request (tracked via a map) and wether \(\pi_D \) is valid. If both conditions hold, it registers \( C \) in the contract state (within a Merkle Tree) and emits \( \text{BootstrapInit}(C, pk_D) \)
\end{itemize}

\subsubsection{Bootstrapping by the authority (A)}
\begin{itemize}
    \item A fetches \( \text{BootstrapInit}\) events
    \item \( b \xleftarrow{\$} \{0,1\}^{256} \)
    \item \( \hat{C} = \text{hash}(C \, \| \, \text{b}) \), add to local database \(<addr_D, \hat{C}>\)
    \item \( \hat{C}_\text{enc} = \text{enc}(pk_D, \hat{C}) \), \(\hat{C}_\text{enc\_hash}=\text{hash}(\hat{C}_\text{enc} \| \hat{C})\)
    \item \(\pi_A \): proving the correctness of the previous calculations starting with a merkle proof for the committed \( C\)
    \item Send \(\{\pi_A, \hat{C}_\text{enc} ,\hat{C}_\text{enc\_hash} \}\) to the mixer
    \item The mixer checks wether \(\pi_A\) is valid; if so, it registers \(\hat{C}_\text{enc\_hash}\) in the contract state and emits \( \text{BootstrappedData}(\hat{C}_\text{enc} ,\hat{C}_\text{enc\_hash}) \)
\end{itemize}

\subsubsection{Actual deposit with the bootstrapped data (D)}
\begin{itemize}
    \item D fetches \text{BootstrappedData} events
    \item D decrypts \(\hat{C}_\text{enc} \), obtaining \(\hat{C}\), which is then constructed as a single element of a Bloom filter (more details on this and the compliance protocol are presented in \hyperref[sub:compliance_features]{section C})
    \item \(\pi_D'\): proving the knowledge of the preimage of \(\hat{C}\_\text{enc\_hash}\), i.e.: \(\hat{C}\_\text{enc}\) and \(\hat{C}\) that only she was able to decrypt. The correct generation of the Bloom filter starting from \(\hat{C}\) is proven as well.
    \item Encrypts the Bloom filter with \(pk_D\) and proceeds with deposit, attaching it to the calldata.
\end{itemize}

\subsection{Private Onboarding}
\begin{figure}
    \centering
    \includegraphics[width=0.85\linewidth]{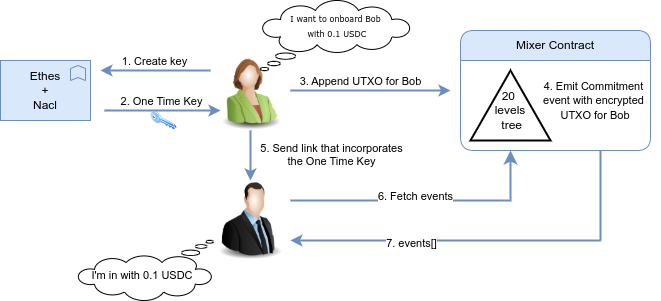}
    \caption{On-boarding procedure in the same mixer as for private transfers.}
    \label{fig:Onboarding}
\end{figure}

A single-tree architecture was used for all commitments—both onboarding and transfer—to be treated as UTXOs within the same Merkle tree structure. Rather than having separate note-based onboarding and UTXO-based transfers, we leverage the creation of a OTK (one-time key) that serves as a temporary encryption key for generating a new UTXO commitment in the shared tree. The advantages of this approach are two-fold:
\begin{itemize}
    \item Architectural simplicity: Instead of maintaining separate systems and circuits for notes and UTXOs, all operations use the same UTXO model and verification mechanisms.
    \item Transaction indistinguishability: This approach mitigates potential onboarding identification by preventing observers from distinguishing between note redemptions and UTXO transfers, as onboarding now appears identical to any other transfer in the system.
\end{itemize} The operational flow, illustrated in Fig. \ref{fig:Onboarding}, is as follows:
\begin{enumerate}
    \item A OTK is randomly generated by Alice according the Nacl encryption scheme \cite{8gwifi_nacl}, consisting in: \( sk_{\text{otk}} \in \mathbb{Z}_p\), \( pk_{\text{otk}} = poseidonHash(sk_{\text{otk}}) \), \( enck_{\text{otk}} = getEncryptionPublicKey(sk_{\text{otk}}.slice(2)) \).
    \item The mixer contract emits an event carrying the UTXO encrypted using \( enck_{\text{otk}}\).
    \item An invitation link containing the OTK is transmitted to Bob.
    \item Upon accessing the link, Bob creates his initially-empty wallet (i.e., adds his pubkey to the onchain registry) and retrieves all events from the mixer
    contract.
    \item Bob's wallet identifies relevant commitment events by attempting decryption with the received \(sk_{\text{otk}} \).
    \\
\end{enumerate} A key aspect of this approach is that Alice, rather than Bob, initiates the UTXO creation and nullifier emission. This architectural choice introduces a trust assumption: Bob must rely on Alice's integrity since she generates and possesses the OTK, granting her theoretical control over initial UTXO. This is mitigated by Bob automatically spending the initial UTXO to a new output controlled by himself, immediately after being onboarded; this would ensure that it is re-encrypted with his key, granting him sole control over it.

\subsection{Compliance features}
\label{sub:compliance_features}

We introduce a two-level compliance system that provides both proactive screening and retroactive flagging capabilities while preserving the privacy guarantees of the underlying UTXO-based pool.

\subsubsection{Proof of Innocence (POI)}

\begin{figure*}
    \centering
    \includegraphics[width=0.85\linewidth]{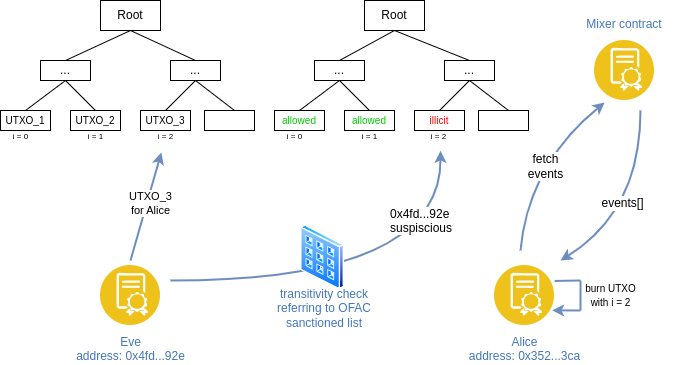}
    \caption{Proof of innocence high-level scheme}
    \label{fig:poi}
\end{figure*}

The first level implements a proof of innocence (POI) mechanism using a parallel Merkle tree structure that maintains compliance states of UTXOs. For a tree of height \( h \), each leaf corresponds to a UTXO in the main pool and contains either "allowed" or "illicit". When a new UTXO is created, its compliance state is determined by:
\begin{itemize}
    \item Direct verification of the depositing address against sanctions lists,
    \item Transitive verification of the address's transaction history up to \( n \) hops, tracing fund flows and checking for sanctioned addresses (see \ref{sec:prevention}), 
    \item Inheritance of compliance states from parent UTXOs in internal transfers.
\end{itemize}
\textbf{Funds withdrawal:} To withdraw and exit the mixer, users must provide zero-knowledge proofs \( \pi_{\text{poi}} \) demonstrating that all input UTXOs reference "allowed" leaves in the parallel tree. The proof statement is: 
\begin{align*}
    & \forall i \in [1,n_{\text{Ins}}]: \\
    & \text{Let } c_i = \text{Poseidon}(a_i, pk_i, b_i) \\
    & \text{where } pk_i = \text{DerivePublicKey}(sk_i) \\
    & \text{If } a_i \neq 0: \\
    & \quad \text{ValidPath}(c_i, \pi_i, \text{idx}_i, R) = 1 
\end{align*}
\label{eq:poi_statement} where Poseidon is a zk-friendly hash function \cite{grassi2021poseidon}, \( c_i \) is the commitment, \( a_i \) is the amount, \( pk_i \) and \( sk_i \) are the public and private keys respectively, \( b_i \) is the blinding factor, \( \pi_i \) is the Merkle path, \( \text{idx}_i \) are the path indices, and \( R \) is the root of the parallel tree.

As shown in Fig. \ref{fig:poi}, a user receiving an "illicit" UTXO can choose to burn it to avoid becoming implicated, relinquishing any control over it. Alternative mechanisms include automatic return to sender, authority seizure or fair redistribution to honest participants via the treasury. These processes are enforceable at the application level, preventing denial-of-service attacks by malicious users attempting to pollute the pool with flagged UTXOs, as illicit funds are automatically subject to these consequences. Future developments could incorporate rate-limiting nullifiers based on Shamir secret sharing \cite{wang2022dske}, where repeated spam attempts would lead to secret reconstruction, effectively de-anonymizing the malicious actor.

\subsubsection{Ancestral Commitment Compliance (ACC)}

\begin{figure*}
    \centering
    \includegraphics[width=0.85\linewidth]{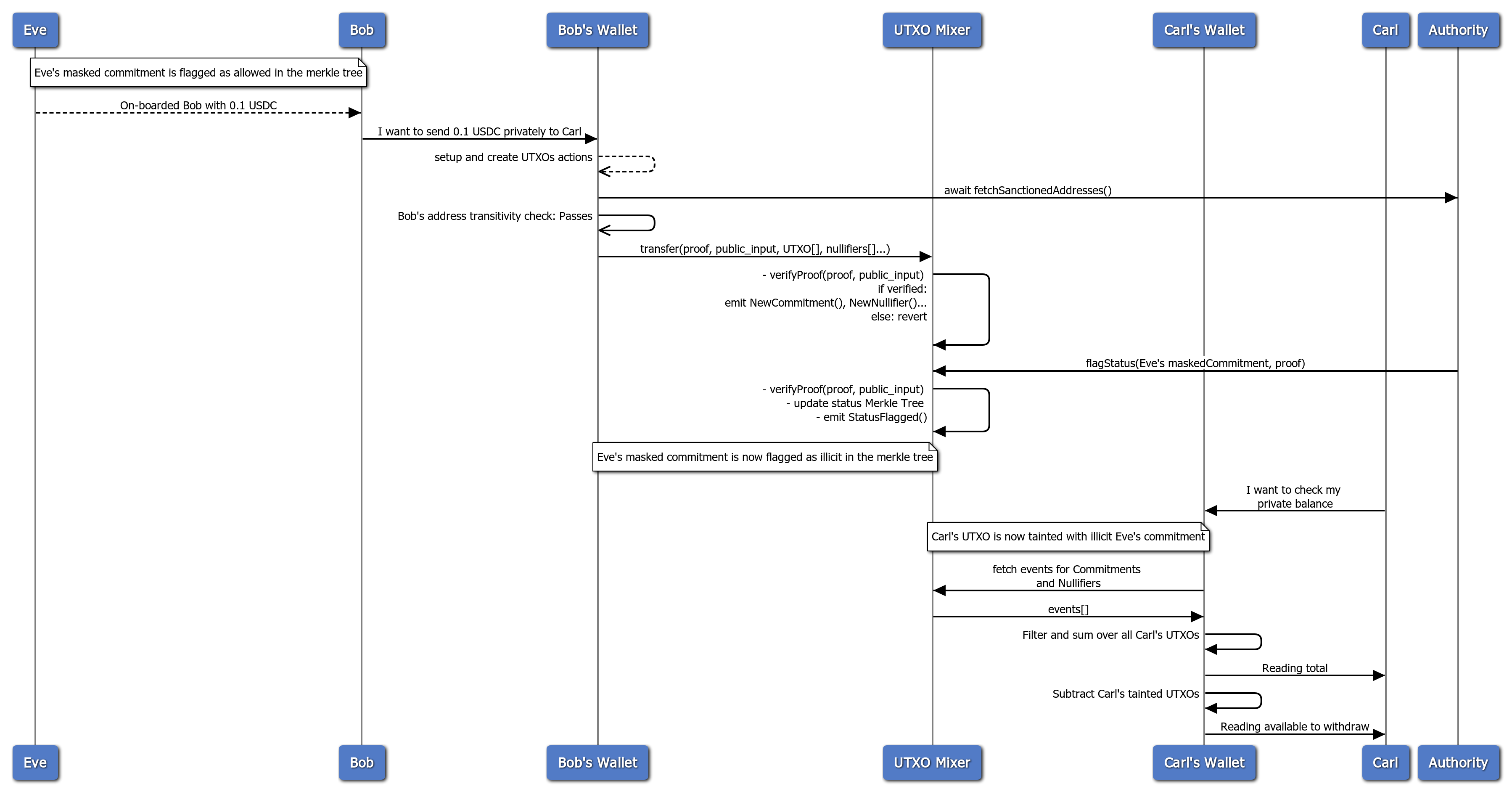}
    \caption{Ancestral Commitment Compliance flow}
    \label{fig:Acenstral_commit}
\end{figure*}

The second level enables ancestral compliance flagging through a chain state propagation mechanism. This can be achieved by allowing an authorized party (e.g. AML officers) to update a separate sparse Merkle tree (SMT) anytime a deposit commitment $C$ becomes illicit. Since deposits in the mixer are indeed visible and can be tracked by anyone, the authorized party keeps an off-chain list of these commitments with their masked versions $\hat{C}$, namely a nested commitment where the blinding factor for hiding is generated by the authority. This masking enables unlinkability - as users learn $\hat{C}$ but not the original $C$ of the received UTXOs, ensuring that the illicit status propagates confidentially through the UTXO chain. 
To refrain from ever-inflating memory allocations, especially after the \textit{coin merge} of UTXOs with long history of $\hat{C}$, our wallet application leverages a bloom filter data structure. A bloom filter is a fixed-size probabilistic data structure designed to test whether an element is a member of a set \cite{bloom1970space}. Each element $\hat{C}$ in the set $S$ is processed by $k$ hash functions, each setting a bit to '1' in an $m$-bit array. Membership queries check these $k$ positions - if any bit is '0', the element is definitely not in $S$, while if all bits are '1', the element is likely present with a configurable false positive rate. Thus, for \( k = \ln(2) \cdot \frac{m}{n} \) (where \( n = |S| \)), the false positive probability is minimized at \( \frac{1}{2^k} \). A sequence diagram of the ancestral commitment compliance flow is described in fig. \ref{fig:Acenstral_commit}. \textbf{Authority Actions:} To update the onchain SMT, authorities must provide zero-knowledge proofs (\(\pi_{\text{mask}}\)) demonstrating that the masked commitment is derived from a valid original commitment that exists in the mixer's Merkle tree to enforce integrity. The proof statement is:
\label{eq:masked_statement}
    \begin{align*}
    & \text{Let } \hat{C} = \text{Poseidon}(C, b), \\
    & \text{where:} \\
    & \quad \hat{C} \text{ is the masked commitment (public input)}, \\
    & \quad C \text{ is the original commitment (private input)}, \\
    & \quad b \text{ is the blinding factor (private input)}. \\
    & \text{For mixer MT root } R_{\text{mixer}} \text{ and Merkle proof } \pi_{\text{mixer}}: \\
    & \quad \text{ValidPath}(C, \pi_{\text{mixer}}, \text{idx}, R_{\text{mixer}}) = 1 \\
    & \text{Assert } \hat{C} = \text{masked commitment}.
    \end{align*}
\label{eq:maskedstatement} 
\\
where \( R_{\text{mixer}} \) is the public root of the mixer's tree, \( \pi_{\text{mixer}} \) is the corresponding Merkle proof for commitment \( C \), and \( \text{idx} \) is the leaf index (private input). When calling the insert function on the SMT contract, the proof \(\pi_{\text{mask}}\) will be atomically verified.

\textbf{Internal Transfer:} To spend their funds, users must decrypt the chain state field of their input UTXOs and fetch \texttt{StatusFlagged} events to check for updates on masked commitments. Before calling \texttt{transact()} via the relayer contract, the user must encrypt the chain state with the recipient's public key and generate a proof of ancestral compliance (\(\pi_{\text{acc}}\)) using the latest SMT root. This proof demonstrates two critical properties: first, that the bloom filter $B_1$ correctly represents the union of all parent UTXOs' chain states being spent; and second, that no flagged ancestral commitment $\hat{C}$ contained in the SMT is a member of this merged bloom filter. Since more than one flagging event might occur during in-between internal transfers, this translates to: \(\forall \hat{C} \in \text{SMT}, \hat{C} \notin S\). The proof (\(\pi_{\text{acc}}\)) statement therefore is:
\begin{align*}
    & \text{Let } B_1[n], B_2[n] \text{ be bit arrays s.t.:} \\
    & \quad B_1[i], B_2[i] \in \{0,1\} \text{ } \forall i \in [0,n-1] \\
    & \text{Assert } B_1 = \bigcup_{j \in \text{inputs}} B_{\text{parent}_j} \text{ (via bitwise OR)} \\
    & \text{For SMT root } R \text{ and Merkle proof } \pi: \\
    & \quad \text{ValidPath}(\hat{C}, B_2, \pi, R) = 1 \\
    & \text{For bloom filter parameter } k: \\
    & \quad \sum_{i=0}^{n-1} (B_1[i] \cdot B_2[i]) \neq k
\end{align*}
where \( B_1 \) is the merged bloom filter representing the union of all parent UTXOs' chain states (computed as the bitwise OR of individual bloom filters), \( B_{\text{parent}_j} \) represents the chain state of the $j$-th input UTXO, \( B_2 \) is the private bit array encoding only the masked commitment \( \hat{C} \), \( k \) is the number of hash functions in the bloom filter, \( R \) is the public root of the SMT containing valid masked commitments, and \( \pi \) is the corresponding Merkle proof, similarly to statement~\ref{eq:poi_statement}. The circuit first verifies the correct computation of the union operation, ensuring that \( B_1[i] = 1 \) if and only if at least one parent UTXO has that bit set. A summation not equal to $k$ would indicate certain exclusion, hence ancestral compliance, because not all bits would be set in the intersection of $B_1$ and $B_2$.

In this scenario, the authority sets the bloom filter parameters and manages the SMT updates. They handle the key-value pairs in the SMT. The key is the masked commitment (in our implementation 32 bytes long), while the value is the Poseidon hash of the fixed-size bit array. The current wallet implementation assumes an honest authority, by which a specific bloom filter construction with SMT key-value pairs is performed consistently. In future versions, statement~\ref{eq:masked_statement} could be extended to verify the bloom filter construction. This extension would prove that an SMT key-value pair was correctly derived using the public parameters \( k \) (hash functions) and m-bit array length.

\section{Evaluation}
A complete flow, from onboarding to private transfer, was tested on Base Sepolia testnet. The bootstrapping overhead in the deposit phase (Section~\ref{sec:deposit}) was not yet evaluated at this stage of the implementation. For onboarding, the only visible trace in the logs is the bundler calling handleOps() on the Entry Point contract \cite{erc4337_architecture_2024}, with only the Relayer address being leaked. For transfers, no actual token movements are visible on-chain - only anonymized event logs appear, indicating that UTXOs were created without revealing their owners or amounts. Notably, each UTXO, containing its entire chain state encoded as bloom filter elements, can be decrypted only by the recipient. We used a bloom filter of size $2^{14}$ bits with $k=2$ hash functions. With these parameters, according to $p = (1 - e^{-\frac{kn}{m-1}})^k$, the false positive rate remains below $5\%$ until $\sim$1600 elements. This provides a good starting point for balancing space efficiency and false positive rate for typical usage scenarios (see Section~\ref{Bloom filter security} for some security considerations).

\subsection{Gas and fees considerations}

Analysis of gas consumption and associated fees shows relatively low transaction costs across all features, with most operations requiring less than 0.01 usd at current rates on the Base network. Table \ref{tab:gas_fees} shows that individual operations like insertIntoPoolUsers consume approximately 84k gas units, while more complex operations such as callWithdraw require up to 430k gas units. The ancestral commitment compliance feature (\( \pi_{\text{acc}} \)) increases transaction costs by a factor of 1.6x compared to proof of innocence alone, primarily due to additional SMT operations and bloom filter chain state manipulations. All these transaction costs bear the overhead of erc-4337 multiple calls and proof verification costs. The solution optimizes gas usage with its userOp ECDSA signature verification mechanism, which is more cost-efficient compared to passkey-based approaches such as that of Daimo wallet \cite{daimo}. However, to make the implementation of a sponsorship model through a paymaster system particularly viable, these costs would need to be further reduced in the future.

\begin{table}[htbp]
\centering
\caption{Gas units and fees}
\resizebox{\columnwidth}{!}{%
\begin{tabular}{|c|c|c|c|c|c|}
\hline
\textbf{Operation} & \textbf{Gas \( \pi_{\text{poi}} \)} & \textbf{Gas \( \pi_{\text{acc}} \)} & \textbf{gWei \( \pi_{\text{poi}} \)} & \textbf{gWei \( \pi_{\text{acc}} \)} & \textbf{usd} \\
\hline
insertIntoPoolUsers & 844001 & 836405 & 84 & 84 & $<$0.01 \\
callDeposit & 1509178 & 2528502 & 151 & 253 & $<$0.01 \\
callTransact & 1716190 & 2842161 & 172 & 285 & $<$0.01 \\
callWithdraw & 1700950 & 2957810 & 170 & 295 & $<$0.01 \\
\hline
\end{tabular}%
}
\label{tab:gas_fees}
\end{table}

\subsection{Computational cost analysis}
The following results show the temporal costs for each feature, including SNARK proof generation and transitivity checks, tested on an ACER Aspire 5 laptop with an Intel Core i7 and NVIDIA GeForce RTX.

Tables \ref{tab:cost_analysis_n2} and \ref{tab:time_cost_analysis_n3} show that most operations complete within reasonable timeframes, with SNARK proof generation consistently efficient at ~2 seconds. The two compliance features add a proving time overhead of around 0.5 and 1.5 sec respectively for withdrawing with proof \( \pi_{\text{poi}} \) and transferring with proof \( \pi_{\text{acc}} \). We tested one single ancestral commitment flagged as illicit, but overhead would scale linearly with additional flagging events. However, this can be optimized through event caching mechanisms that only fetch and verify newly flagged commitments since the last check.
The longer durations in operations like join via invite code (\(\sim 26\,\text{s} \))
 are due to executing both insertIntoPoolUsers and callDeposit sequentially, with necessary mining delays between them to prevent nonce collisions. Similarly, onboarding via link (\(\sim 15\,\text{s} \)) requires both insertIntoPoolUsers and insertIntoEncryptedData with appropriate delays. The transitivity check demonstrates significant computational overhead, with each additional hop increasing processing time by roughly 7x, suggesting careful consideration is needed when implementing deeper transaction history checks.

\begin{table}[htbp]
\centering
\caption{Cost analysis considering \( n=2 \) in the transitivity check}
\resizebox{\columnwidth}{!}{%
\begin{tabular}{|c|c|c|c|}
\hline
\textbf{Action} & \textbf{SNARK proof (ms)} & \textbf{Transitivity check (n=2)} & \textbf{Overall time (ms)} \\
\hline
Join via invite code & 1334.72 & X & 25941.17 \\
Onboard someone & 2097.78 & X & 7200.74 \\
Onboard via link & X & X & 14486.89 \\
Fund the wallet & 2064.56 & 1193.75 & 26226.83 \\
Transfer privately & 1999.40 & 1730.96 & 10792.50 \\
Transfer privately \( \pi_{\text{acc}} \) & 1999.40 + 1536.33 & 1730.96 & 12328.83 \\
Withdraw \( \pi_{\text{poi}} \) & 2076.77 + 727.88 & X & 11847.13 \\
\hline
\end{tabular}%
}
\label{tab:cost_analysis_n2}
\end{table}

\begin{table}[htbp]
\centering
\caption{Time cost analysis considering \( n=3 \) in the transitivity check}
\resizebox{\columnwidth}{!}{%
\begin{tabular}{|c|c|c|c|}
\hline
\textbf{Action} & \textbf{SNARK proof (ms)} & \textbf{Transitivity check (n=3)} & \textbf{Overall time (ms)} \\
\hline
Join via invite code & 2156.48 & X & 28071.54 \\
Onboard someone & 672.06 & X & 5967.50 \\
Onboard via link & X & X & 16499.45 \\
Fund the wallet & 702.58 & 7332.75 & 26540.74 \\
Transfer privately & 2123.27 & 7817.52 & 15983.88 \\
Transfer privately \( \pi_{\text{acc}} \) & 2123.27 + 1574.54 & 7817.52 & 17558.42 \\
Withdraw \( \pi_{\text{poi}} \) & 2017.69 + 542.55 & X & 11669.14 \\
\hline
\end{tabular}%
}
\label{tab:time_cost_analysis_n3}
\end{table}

\subsection{AML Considerations and Remediation Mechanics} \label{AML Considerations}

The tension between privacy preservation and regulatory compliance onchain has historically been positioned as a binary choice \cite{Corp2023Case}. Our ancestral commitment compliance approach attempts to bridge this gap by providing participants with probabilistic knowledge about the provenance of their funds — certain when legitimate, probable when illicit. This shifts AML considerations from rigid classification to a more nuanced risk-based framework, which better reflects the complexities of financial systems, where absolute certainty is rare and risk mitigation should be adaptive \cite{FATF2023}.

If a UTXO is found to contain ancestrally tainted funds, remediation options vary in complexity and impact. One approach is voluntary burning, which removes the funds but unfairly penalizes recipients. Alternatively, tainted UTXOs could be returned to the sender, though this only shifts the problem backward. A more structured solution involves directing tainted funds to an authority or community treasury, enabling governance-based redistribution.

A possible "cleansing" mechanism could involve a rehabilitation protocol where users prove good-faith transactions over time. Inspired by traditional finance’s amnesty or compliance-based reintegration \cite{fatf_taxamnesty_2025}, this could involve placing tainted funds in escrow and releasing them upon meeting predefined legitimacy criteria.

\subsubsection{The Fungibility Paradox and Practical Enforcement}

Our system introduces what might be called a "fungibility paradox" in digital assets. While blockchain tokens are technically fungible at the protocol level (except for non-fungible tokens like ERC-721), our ancestral tracing creates a layer of "practical non-fungibility" based on historical provenance. 
In traditional finance, a banknote that passed through illicit channels before entering general circulation remains spendable by downstream recipients under good faith acquisition doctrines. The cost of tracing and reclaiming every tainted physical note would be prohibitive, but digital assets follow a distinct logic. With our solution, the marginal cost of identifying tainted funds approaches zero. This leads to some practical and ethical considerations:

\begin{itemize}
    \item \textbf{Temporal decay of responsibility}: Should a UTXO's "illicit" status decay over time or transaction hops, similar to statutes of limitation in traditional law?
    
    \item \textbf{De minimis thresholds}: Should trace amounts of tainted funds (e.g., below 1\% of a UTXO's value) trigger the same responses as predominantly tainted UTXOs?
    
    \item \textbf{Innocent holder protection}: Can algorithmic markers distinguish between knowing participants in illicit transactions and innocent recipients?
\end{itemize} These considerations suggest implementing graduated responses rather than binary classifications, potentially assigning each UTXO a continuous "compliance score" that determines appropriate remediation pathways.

\subsection{Security and Privacy considerations}

We discuss the security and privacy strengths and weaknesses across these dimensions: Proof system soundness, Collision resistance, Bloom filter privacy and compliance guarantees and overall wallet architecture security. 

\subsubsection{Proof system soundness} The application relies on Groth16, which provides zk-SNARKs with constant size and efficient verification \cite{cryptoeprint:2016/260}. While Groth16 requires a trusted setup, this is mitigated by using the established setup from Tornado Cash Nova for the base JoinSplit circuit which is here used for the additional proof of innocence and ancestral commitment circuits. The commitment scheme $c = \sum_i w_i \cdot b_i$ - where $w_i$ are the private witness values and $b_i$ are points on the BN254 curve, relies on the hardness of the discrete logarithm problem. Advances in the Special Tower Number Field Sieve attack have reduced its security from from 128 bits to (\(\sim 102\,\text{bits} \)) \cite{chaliasos2024sok}, lowering the soundness of each proof to a forgery probability of $2^{-102}$, below the recommended 128-bit for cryptographic applications, though this is arguably less concerning as even some high-volume layer 2 rollups operate with provers below 128 bits. Finally, the vulnerability often associated with incorrect implementations of the Fiat-Shamir transformation \cite{dao2023weak}, does not affect our standard Groth16-based solution.
\subsubsection{Proof Non-Malleability}
Our system leverages the inherent non-malleability properties in \cite{cryptoeprint:2016/260} to ensure that a valid proof cannot be transformed into a different valid proof for the same statement without knowledge of the witness. Furthermore, our implementation binds each proof to its specific transaction context via a hash of critical parameters:
\begin{equation*}
\text{txContext} = \text{Hash}\Biggl(
\begin{aligned}
    &R \parallel \text{inputNullifiers} \parallel \\
    &\text{outputCommitments} \parallel \text{publicAmount}
\end{aligned}
\Biggr)
\end{equation*} where $R$ is the Merkle tree root. This binding, together with the use of ECDSA signatures to secure the transaction data, ensures that any attempt to modify a proof (e.g., by altering the recipient address) invalidates the proof. As a result, replay attacks and proof pollution are effectively prevented.

\subsubsection{Collision resistance} 
In our chain state propagation design, when the authority flags a masked commitment as illicit, they have computed: $\hat{C} = \text{Poseidon}(C, b)$
where $C$ is the original commitment and $b$ is a blinding factor. Furthermore, the authority uses hash functions to compute the indices for the bytes32 element to be check against the bloom filter. In either case, for Poseidon $H: {0,1}^* \rightarrow {0,1}^n$, collision resistance means an adversary cannot efficiently find $x_1, x_2$ where $x_1 \neq x_2$ such that $H(x_1) = H(x_2)$. Collision resistance is achieved through Poseidon construction based on the non-linear permutation using S-boxes defined as $S(x) = x^d$ over the prime field $\mathbb{F}_p$ where $d$ is $5$ and $\mathbb{F}_p$ is 254-bit, the linear MDS matrix multiplication and round constants \cite{grassi2021poseidon}. Specifically, for bloom filter indices, given inputs $x_1 \neq x_2$, it should be computationally infeasible to find collisions in $\text{Poseidon}(x_1) \bmod m = \text{Poseidon}(x_2) \bmod m$. This ensures the system maintains 128-bit security against collision attacks.

\subsubsection{Bloom filter security} \label{Bloom filter security} We propose strategies to protect our bloom-filter system from two attack vectors: 

\begin{enumerate}[label=\roman*.]
    \item Chain state pollution with repeated coin merges that could inflate the false positive rate.
    \label{pollution}
    \item Input manipulation where a malicious authority could carefully select inputs to maximize hash collisions \cite{vos2024insecurity}. 
    \label{manipulation}
\end{enumerate}

\ref{pollution} For a bloom filter with $m$ bits and $k$ hash functions, the probability of false positives after $n$ merges can be expressed as:
\[ P_{fp}(n) = \left(1 - e^{-\frac{kn}{m}}\right)^k \cdot P_{collision} \]
where $P_{collision}$ is negligible under standard cryptographic assumptions (see above). To prevent excessive false positives from repeated merges, we propose a rate limiting $R_{max}$ at the smart contract level for a block time period $t$:
\[ R_{max}(t) = \frac{m}{k} \ln\left(\frac{1}{\tau(t)}\right) \]
where $\tau(t)$ is an adaptive threshold that becomes more stringent as transaction volume increases within a time window $t$:
\[ \tau(t) = \tau_{base} \cdot e^{-\alpha V(t)} \]
with $V(t)$ representing the transaction volume and $\alpha$ a tuning parameter. Future implementations could require users to create fresh commitments with new bloom filters when capacity thresholds are reached, preserving the cryptographic proof chain between old and new states.

\vspace{10pt}

\ref{manipulation} The authority only adds single masked commitments $\hat{C}$ into the SMT, with each $\hat{C}$ encoded as one element in an empty bloom filter. While the authority could theoretically try to select $\hat{C}$ values that maximize collisions in the bloom filter, this attack vector is severely limited since chain states are encrypted with recipient keys. Even in case of collusion with a recipient, visibility would be restricted to only those chain states involved in transactions with the colluding party. Therefore, while we could maintain a per-epoch limit on SMT insertions as a safeguard (i.e., $\text{inserted}/\text{epoch} < \frac{m}{2k}$), this would be primarily a precautionary measure against DoS attacks rather than a strict security requirement for the system's privacy guarantees.

Overall, our design mitigates transaction graph manipulation attacks, funds laundering attempts. Since every UTXO inherits the union of its parent UTXOs' bloom filters, and flagged commitments are permanently recorded in the authority's SMT, no sequence of transfers can remove a taint once applied. Should temporal decay or de minimis thresholds be implemented in the future (see Section \ref{AML Considerations}), a different threat model would arise requiring additional safeguards against both attack vectors above.

\subsubsection{Smart wallet security} 

The wallet architecture implements standard security measures to protect against smart contract vulnerabilities that often lead to unauthorized operations, frontrunning, or exploiting gasless transactions.
In the EntryPoint contract, a replay protection mechanism employs a double-layer nonce structure for userOp, defined as \( \text{nonce} = (\text{key} \ll 64) | \text{seq} \), where \(\text{key}\) is a 192-bit value and \(\text{seq}\) is a 64-bit sequence number. Validation is performed with \( \text{nonceSequenceNumber}[\text{sender}][\text{key}]++ == \text{seq} \), ensuring each nonce is used only once as the sequence number increments immediately after validation. 

For protection against reentrancy attacks, the account contract inherits from OpenZeppelin's ReentrancyGuard and strictly adheres to the checks-effects-interactions pattern. 
The relayer architecture provides additional security through other mechanisms: The bundler's handleOps() function ensures atomic execution guarantees, while mandatory signature verification precedes any state-changing operations. To prevent DoS attacks, strict gas limits are enforced on userOp execution.

\section{Conclusions}
Regulatory alignment in blockchain applications remains an ever-present challenge. We have proposed a smart wallet implementation that embeds privacy and compliance features by design. The first compliance feature is represented by proof of innocence generation upon withdrawal, where compliance states of input UTXOs are checked against a pool of sanctioned addresses and their transaction histories maintained in a parallel Merkle tree. We further propose a mechanism for managing UTXO chain states through fixed-length bloom filter bit arrays. In this approach, for transactions consuming multiple UTXOs, newly created ones inherit a composite state through a union operation of predecessor states. The probabilistic nature of bloom filter checks represents a key trade-off in determining fund legitimacy. This architecture opens up new possibilities for implementing nuanced compliance policies, especially for AML requirements, where blockchain operations can be either completely "clean" or potentially suspect with quantifiable degrees of uncertainty. While our current implementation assumes a honest authority, verifying bloom filter construction in zero knowledge would reduce this trust requirement in future versions. We expect subsequent work to explore this trade-off more deeply, particularly examining how authorized parties can effectively regulate privacy-preserving applications with fair oversight and autonomy.

\bibliographystyle{IEEEtran}
\bibliography{References}

\begin{thebibliography}{10}
\providecommand{\url}[1]{#1}
\csname url@samestyle\endcsname
\providecommand{\newblock}{\relax}
\providecommand{\bibinfo}[2]{#2}
\providecommand{\BIBentrySTDinterwordspacing}{\spaceskip=0pt\relax}
\providecommand{\BIBentryALTinterwordstretchfactor}{4}
\providecommand{\BIBentryALTinterwordspacing}{\spaceskip=\fontdimen2\font plus
\BIBentryALTinterwordstretchfactor\fontdimen3\font minus \fontdimen4\font\relax}
\providecommand{\BIBforeignlanguage}[2]{{%
\expandafter\ifx\csname l@#1\endcsname\relax
\typeout{** WARNING: IEEEtran.bst: No hyphenation pattern has been}%
\typeout{** loaded for the language `#1'. Using the pattern for}%
\typeout{** the default language instead.}%
\else
\language=\csname l@#1\endcsname
\fi
#2}}
\providecommand{\BIBdecl}{\relax}
\BIBdecl

\bibitem{kumar2024blockchain}
J.~Kumar, G.~Rani, M.~Rani, and V.~Rani, ``Blockchain technology adoption and its impact on sme performance: insights for entrepreneurs and policymakers,'' \emph{Journal of Enterprising Communities: People and Places in the Global Economy}, 2024.

\bibitem{usd2022}
\BIBentryALTinterwordspacing
{United States Department of the Treasury}, ``{U.S. Treasury Sanctions Notorious Virtual Currency Mixer Tornado Cash},'' 2022, accessed: 2024-12-21. [Online]. Available: \url{https://home.treasury.gov/news/press-releases/jy0916}
\BIBentrySTDinterwordspacing

\bibitem{wu2023towards}
J.~Wu, D.~Lin, Q.~Fu, S.~Yang, T.~Chen, Z.~Zheng, and B.~Song, ``Towards understanding asset flows in crypto money laundering through the lenses of ethereum heists,'' \emph{IEEE Transactions on Information Forensics and Security}, 2023.

\bibitem{schellinger2022yes}
B.~Schellinger, F.~V{\"o}lter, N.~Urbach, and J.~Sedlmeir, ``Yes, i do: Marrying blockchain applications with gdpr,'' \emph{e-government}, vol.~19, p.~22, 2022.

\bibitem{bamberger2022verification}
K.~A. Bamberger, R.~Canetti, S.~Goldwasser, R.~Wexler, and E.~J. Zimmerman, ``Verification dilemmas in law and the promise of zero-knowledge proofs,'' \emph{Berkeley Tech. LJ}, vol.~37, p.~1, 2022.

\bibitem{buterin2024blockchain}
V.~Buterin, J.~Illum, M.~Nadler, F.~Sch{\"a}r, and A.~Soleimani, ``Blockchain privacy and regulatory compliance: Towards a practical equilibrium,'' \emph{Blockchain: Research and Applications}, vol.~5, no.~1, p. 100176, 2024.

\bibitem{spagnuolo2014bitiodine}
M.~Spagnuolo, F.~Maggi, and S.~Zanero, ``Bitiodine: Extracting intelligence from the bitcoin network,'' in \emph{Financial Cryptography and Data Security: 18th International Conference, FC 2014, Christ Church, Barbados, March 3-7, 2014, Revised Selected Papers 18}.\hskip 1em plus 0.5em minus 0.4em\relax Springer, 2014, pp. 457--468.

\bibitem{wahrstatter2024basesap}
A.~Wahrst{\"a}tter, M.~Solomon, B.~DiFrancesco, V.~Buterin, and D.~Svetinovic, ``Basesap: Modular stealth address protocol for programmable blockchains,'' \emph{IEEE Transactions on Information Forensics and Security}, 2024.

\bibitem{barbereau2023beyond}
T.~Barbereau, E.~Ermolaev, M.~Brennecke, E.~Hartwich, and J.~Sedlmeir, ``Beyond a fistful of tumblers: Toward a taxonomy of ethereum-based mixers,'' 2023.

\bibitem{9520375}
X.~Sun, F.~R. Yu, P.~Zhang, Z.~Sun, W.~Xie, and X.~Peng, ``A survey on zero-knowledge proof in blockchain,'' \emph{IEEE Network}, vol.~35, no.~4, pp. 198--205, 2021.

\bibitem{maxwell2013coinjoin}
G.~Maxwell, ``Coinjoin: Bitcoin privacy for the real world, 2013,'' \emph{URl: https://bitcointalk. org/index. php}, 2013.

\bibitem{gabbay2021algebras}
M.~J. Gabbay, ``Algebras of utxo blockchains,'' \emph{Mathematical Structures in Computer Science}, vol.~31, no.~9, pp. 1034--1089, 2021.

\bibitem{jivanyan2019lelantus}
A.~Jivanyan, ``Lelantus: Towards confidentiality and anonymity of blockchain transactions from standard assumptions.'' \emph{IACR Cryptol. ePrint Arch.}, vol. 2019, p. 373, 2019.

\bibitem{tornado_cash_docs}
{Tornado Cash}, ``Tornado cash documentation,'' \url{https://github.com/tornadocash/docs}, n.d., accessed: 2024-11-08.

\bibitem{railgun_wiki}
``Railgun wiki,'' \url{https://docs.railgun.org/wiki}, n.d., \url{https://docs.railgun.org/wiki}.

\bibitem{burleson2022privacy}
J.~Burleson, M.~Korver, and D.~Boneh, ``Privacy-protecting regulatory solutions using zero-knowledge proofs,'' 2022.

\bibitem{beal2023derecho}
J.~Beal and B.~Fisch, ``Derecho: Privacy pools with proof-carrying disclosures,'' \emph{Cryptology ePrint Archive}, 2023.

\bibitem{railgun_2024}
{Railgun}, ``Railgun,'' \url{https://www.railgun.org/}, 2024.

\bibitem{ofac_sdn_list_2024}
{Office of Foreign Assets Control}, ``Specially designated nationals and blocked persons list (sdn) human readable lists,'' \url{https://ofac.treasury.gov/specially-designated-nationals-and-blocked-persons-listsdn-human-readable-lists}, 2024, u.S. Department of the Treasury.

\bibitem{popov2024blockchain}
V.~Popov, M.~Krupin, A.~Gross, and G.~Koreli, ``Blockchain privacy and self-regulatory compliance: Methods and applications,'' \emph{Available at SSRN 4787693}, 2024.

\bibitem{hinkal_docs}
H.~Team, ``Hinkal documentation,'' \url{https://hinkal-team.gitbook.io/hinkal}, accessed: 2024-11-08.

\bibitem{ohlhaver2022decentralized}
P.~Ohlhaver, E.~G. Weyl, and V.~Buterin, ``Decentralized society: Finding web3's soul,'' \emph{Available at SSRN 4105763}, 2022.

\bibitem{gulamov2021tornado}
\BIBentryALTinterwordspacing
I.~Gulamov, ``Tornado pool audit,'' Tornado Cash, Tech. Rep., 2021, accessed: 2021. [Online]. Available: \url{https://github.com/tornadocash/tornado-nova/blob/master/resources/Zeropool-Tornado.pool-audit.pdf}
\BIBentrySTDinterwordspacing

\bibitem{erc4337_architecture_2024}
{ERC-4337}, ``Erc-4337 architecture,'' \url{https://www.erc4337.io/docs/understanding-ERC-4337/architecture}, 2024.

\bibitem{8gwifi_nacl}
\BIBentryALTinterwordspacing
{8gwifi.org}, ``Nacl encryption using golang,'' 2018. [Online]. Available: \url{https://8gwifi.org/docs/go-nacl.jsp}
\BIBentrySTDinterwordspacing

\bibitem{grassi2021poseidon}
L.~Grassi, D.~Khovratovich, C.~Rechberger, A.~Roy, and M.~Schofnegger, ``Poseidon: A new hash function for $\{$Zero-Knowledge$\}$ proof systems,'' in \emph{30th USENIX Security Symposium (USENIX Security 21)}, 2021, pp. 519--535.

\bibitem{wang2022dske}
Z.~Wang, O.~Alpos, A.~Kavousi, S.~Y. Chau, D.~V. Le, and C.~Cachin, ``Dske: Digital signature with key extraction,'' \emph{Cryptology ePrint Archive}, 2022.

\bibitem{bloom1970space}
B.~H. Bloom, ``Space/time trade-offs in hash coding with allowable errors,'' \emph{Communications of the ACM}, vol.~13, no.~7, pp. 422--426, 1970.

\bibitem{daimo}
\BIBentryALTinterwordspacing
D.~Contributors, ``Daimo: A modular framework for building ethereum and evm-based infrastructure,'' 2024, accessed: 2024-12-19. [Online]. Available: \url{https://github.com/daimo-eth/daimo}
\BIBentrySTDinterwordspacing

\bibitem{Corp2023Case}
{Corp., Shlomit Azgad-Tromer, CEO and Chief Legal Officer at Sealance}, {Bank, Joey Garcia, Director and Head of Legal \& Regulatory Affairs at Xapo}, and A.~R. S. a.~C. University, Eran~Tromer, ``The {Case} for {On}-{Chain} {Privacy} and {Compliance},'' \emph{Stanford Journal of Blockchain Law \& Policy}, jun 24 2023, https://stanford-jblp.pubpub.org/pub/onchain-privacy-compliance.

\bibitem{FATF2023}
\BIBentryALTinterwordspacing
{Financial Action Task Force}, ``Fatf guidance on the risk-based approach to combating money laundering and terrorist financing: High-level principles and procedures,'' 2023, accessed: 2025-02-26. [Online]. Available: \url{https://www.fatf-gafi.org/en/publications/Fatfrecommendations}
\BIBentrySTDinterwordspacing

\bibitem{fatf_taxamnesty_2025}
\BIBentryALTinterwordspacing
------, ``Tax amnesty and asset repatriation programmes,'' 2025, accessed: 2025-02-26. [Online]. Available: \url{https://www.fatf-gafi.org/en/publications/Fatfrecommendations/Taxamnestyandassetrepatriationprogrammes.html}
\BIBentrySTDinterwordspacing

\bibitem{cryptoeprint:2016/260}
\BIBentryALTinterwordspacing
J.~Groth, ``On the size of pairing-based non-interactive arguments,'' Cryptology {ePrint} Archive, Paper 2016/260, 2016. [Online]. Available: \url{https://eprint.iacr.org/2016/260}
\BIBentrySTDinterwordspacing

\bibitem{chaliasos2024sok}
S.~Chaliasos, J.~Ernstberger, D.~Theodore, D.~Wong, M.~Jahanara, and B.~Livshits, ``Sok: What don't we know? understanding security vulnerabilities in snarks,'' \emph{arXiv preprint arXiv:2402.15293}, 2024.

\bibitem{dao2023weak}
Q.~Dao, J.~Miller, O.~Wright, and P.~Grubbs, ``Weak fiat-shamir attacks on modern proof systems,'' in \emph{2023 IEEE Symposium on Security and Privacy (SP)}.\hskip 1em plus 0.5em minus 0.4em\relax IEEE, 2023, pp. 199--216.

\bibitem{vos2024insecurity}
J.~Vos, J.~van Assen, T.~Koster, E.~A. Markatou, and Z.~Erkin, ``On the insecurity of bloom filter-based private set intersections,'' \emph{Cryptology ePrint Archive}, 2024.

\end{thebibliography}

\end{document}